\documentstyle[epsfig]{aipproc}

\def\msb{\overline{\rm MS}}
\def\almz{\alpha_s(M_Z)} 
\def\lapproxeq{\lower .7ex\hbox{$\;\stackrel{\textstyle
<}{\sim}\;$}}
\def\gapproxeq{\lower .7ex\hbox{$\;\stackrel{\textstyle
>}{\sim}\;$}}
\def\be{\begin{equation}}
\def\ee{\end{equation}}
\def\bea{\begin{eqnarray}}
\def\eea{\end{eqnarray}}

\begin{document}
\begin{flushright}
RAL--TR--97--024
\end{flushright}
\title{Phenomenology of Deep Inelastic Scattering
Structure Functions}

\author{R. G. Roberts}
\address{Rutherford Appleton Laboratory\\
Chilton, Didcot, Oxon OX12 0QX UK
\thanks{Plenary Talk at 5th International Workshop on Deep Inelastic
Scattering and QCD (DIS'97), Chicago, April 1997}
}
\maketitle

\begin{abstract}
I review recent progress in analysing deep inelastic scattering 
structure functions in global analyses. The new ingredients are 
new data and
attempts to incorporate heavy quarks consistently. A new way of
including the resummation of large $\log 1/x$ terms is discussed.
\end{abstract}

\section*{Introduction}
Since we last met in Rome last year there has been progress in our 
understanding of deep inelastic scattering
(DIS) structure functions and the implications of new measurements 
in the context of global
analyses and extraction of parton distribution functions (pdf's). 
 Firstly there have been updates to two
major experiments $-$ the muon-nucleon NMC collaboration at CERN and the 
neutrino-iron CCFR collaboration at Fermilab. The NMC \cite{nmc} has
produced their final numbers for $F_2$ for proton and deuterium targets
at 90, 120, 200 and 280 GeV while the CCFR \cite{ccfr} collaboration has
 implemented new
energy calibrations to its analysis. The latter data are consistent
with a greater value of $\almz$ than the previously reported value.
Finally there have been the first reported measurements
from HERA on the charm structure function $F_2^c$ \cite{charmh1,charmzeus}
which indicate that a substantial fraction of the total $F_2$ at HERA
comes from charm production.

On the theoretical side, progress has been made following
improved treatments of
heavy quark production in DIS. Why is this important? From above we see
that the component of $F_2$ is relatively large and so it is obviously 
important to have a consistent description of this component and to check 
this with boson-gluon fusion mechanisms. Also if we are trying to
understand the small $x$ behaviour of $F_2$ (i.e. BFKL versus GLAP)
then the data force us to study small $Q^2$ i.e. the region where
we move through the charm threshold; so this should be understood.
Finally it is necessary to produce pdf's for charm and bottom flavours
to insert into other processes such as jet production.

Large $x$, high $Q^2$ is where the excitement resides just now. Can we
be precise about the conventional background and estimate the uncertainty
in the cross section around $Q^2 \sim 10^5 GeV^2$, $x \sim 0.45$? I shall
discuss this briefly.

For some time it has been realised that in addition to resumming large
logs in $Q^2$ (GLAP), there may be significant contributions arising
from resumming potentially large logs in $1/x$ (BFKL). The difficulty
is to devise a theoretically consistent procedure of `
marrying' the two in a
practical way that confronts the data and tries to answer the question
whether these data favour or disfavour the inclusion of the $\log 1/x$ 
resummed terms. Here there has been definite progress and I shall 
highlight a recent analysis by Thorne \cite{thorne}.

\section*{Recent global analyses}
The two major providers of pdf's, CTEQ and MRS, continue to update
their global analyses of DIS and other processes. Both are now attempting
to include a theoretically improved treatment of charm and bottom and
these treatments are discussed in the following section. In this section
I will highlight some results emerging from these on-going analyses.
\begin{figure}[hbt] 
\centerline{\epsfig{file=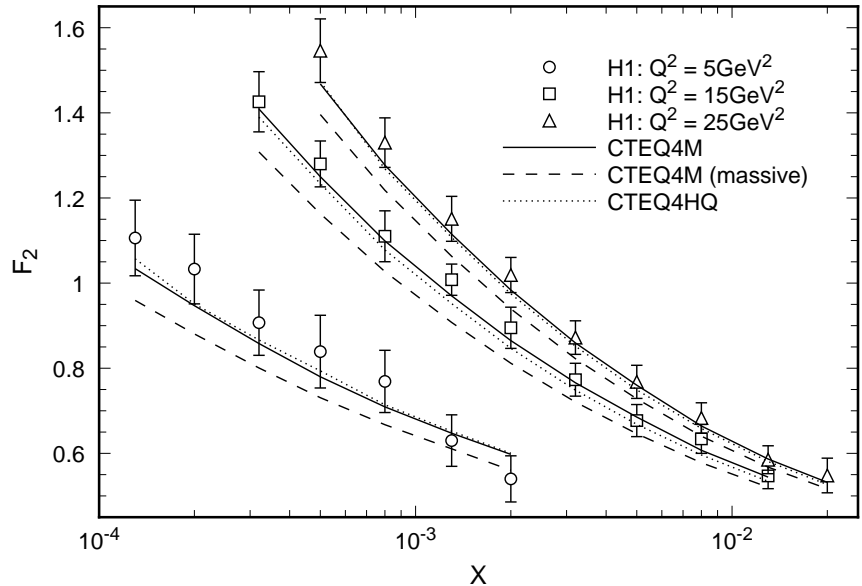,height=3.2in}}
\vspace{10pt}
\caption{Comparison of H1 data\protect\cite{h1f2}
 with calculations using CTEQ4M pdf's
in the original massless scheme (solid line) with the new fit CTEQ4HQ
using the ACOT procedure (dotted line).}
\label{fighera}
\end{figure}
Lai and Tung\cite{lai} demonstrate the effect of including the variable
flavour scheme (VFNS) developed by Aivazis {\it et al.}\cite{acot} (ACOT) 
based on their ealier work\cite{wkt} into the CTEQ analysis. This is to 
be contrasted with the previous treatment in CTEQ where charm evolved
as a massless quark. The effect is illustrated in fig. \ref{fighera}.

\begin{figure}[htb] 
\centerline{\epsfig{file=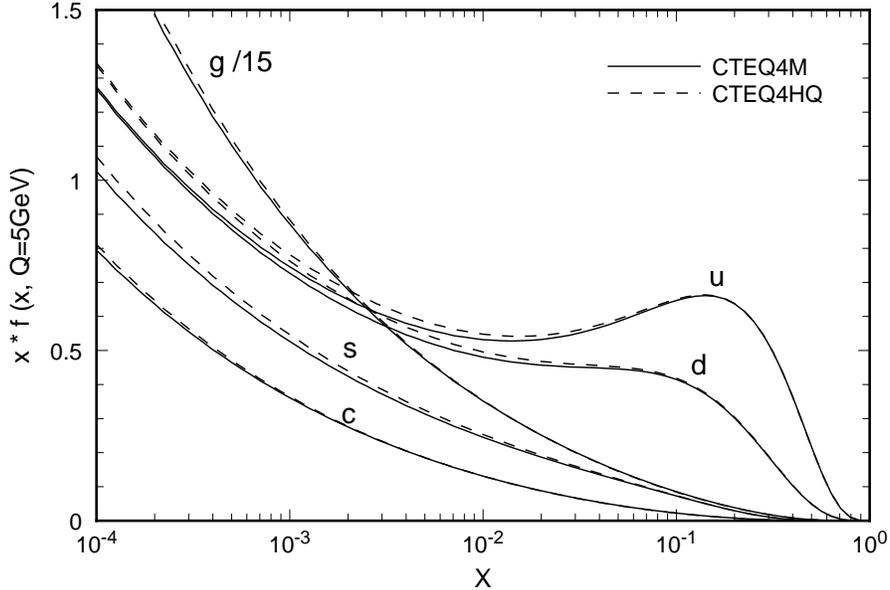,height=3.2in}}
\vspace{10pt}
\caption{Comparison of CTEQ4M and CTEQ4HQ pdf's}
\label{figpdf}
\end{figure}

Only a small change is observed as one changes from CTEQ4M (the massless
evolution) to the improved VFNS procedure which is more noticeable at
small $x$. A comparison of the resulting pdf's at $Q^2=25$ GeV$^2$ is shown 
in fig. \ref{figpdf}
Moreover, the quality of the resulting fit to the full $F_2$ is improved
at low $x$. 

\begin{figure}[htb] 
\centerline{\epsfig{file=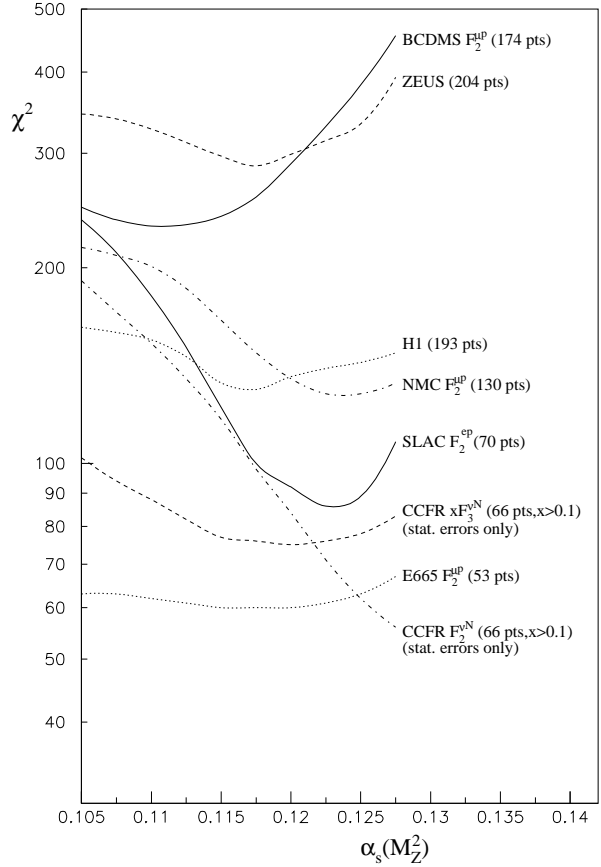,height=5.2in}}
\vspace{10pt}
\caption{Values of $\chi^2$ for various DIS 
datasets\protect\cite{h1f2,zeusf2,bcdms,nmc,slac,e665}
as $\almz$ is varied}
\label{chisq}
\end{figure}

The series of global analyses carried out by MRS has been updated. The new
data from NMC\cite{nmc} and CCFR\cite{ccfr} are included and the improved
treatment of the heavy quark developed by Martin {\it et al.}\cite{mrrs}
is used (see next section). As with the previous MRS analysis\cite{mrsr}
a low value of the starting scale $Q_0^2$ of 1 GeV$^2$ is taken. This is
mainly to reach as small $x$ as possible for the HERA 
data\cite{h1f2,zeusf2} for which a cut at $Q^2 = 1.5$ GeV$^2$ is taken 
while the cut is at 2 GeV$^2$ for other DIS data.  In order to study
the sensitivity to the value of $\almz$, the analysis is repeated at
fixed values of $\almz$ over a wide range. The resulting values of the
individual $\chi^2$ values for the DIS data sets are shown in fig. \ref{chisq}.
From the figure we see (i) that the HERA\cite{h1f2,zeusf2}
 data prefer a value of $\almz$ around 0.12 or so\cite{bf1}, 
(ii) that the updated NMC\cite{nmc} data also prefer this somewhat
larger value and (iii) the re-analysed CCFR\cite{ccfr} $F_2$ data 
{\it strongly} favour a value at the upper end. This leaves 
BCDMS\cite{bcdms} data as the one set which still favours a relatively
low value of $\almz$. But we should recall that these data are very precise
and measure scaling violations in the large $x$ region where the
determination of $\almz$ is insensitive to possible uncertainty of the gluon
distribution.  

\begin{figure}[htb] 
\centerline{\epsfig{file=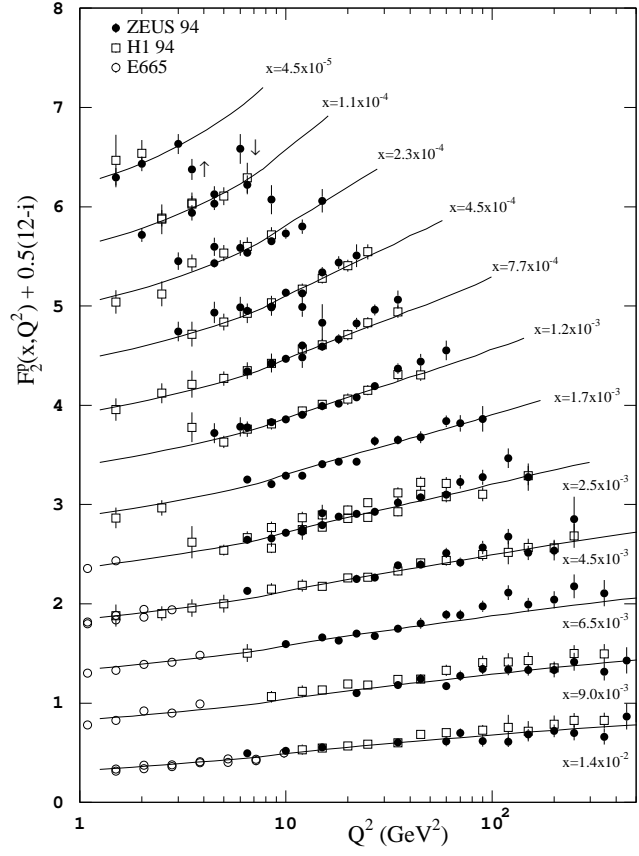,height=5.2in}}
\vspace{10pt}
\caption{Comparison of the new fit using $\almz = 0.118$ with the
small $x$ data from refs\protect\cite{h1f2,zeusf2,e665}.}
\label{smallx}
\end{figure}

\begin{figure}[htb] 
\vspace{-2.5cm}
\centerline{\epsfig{file=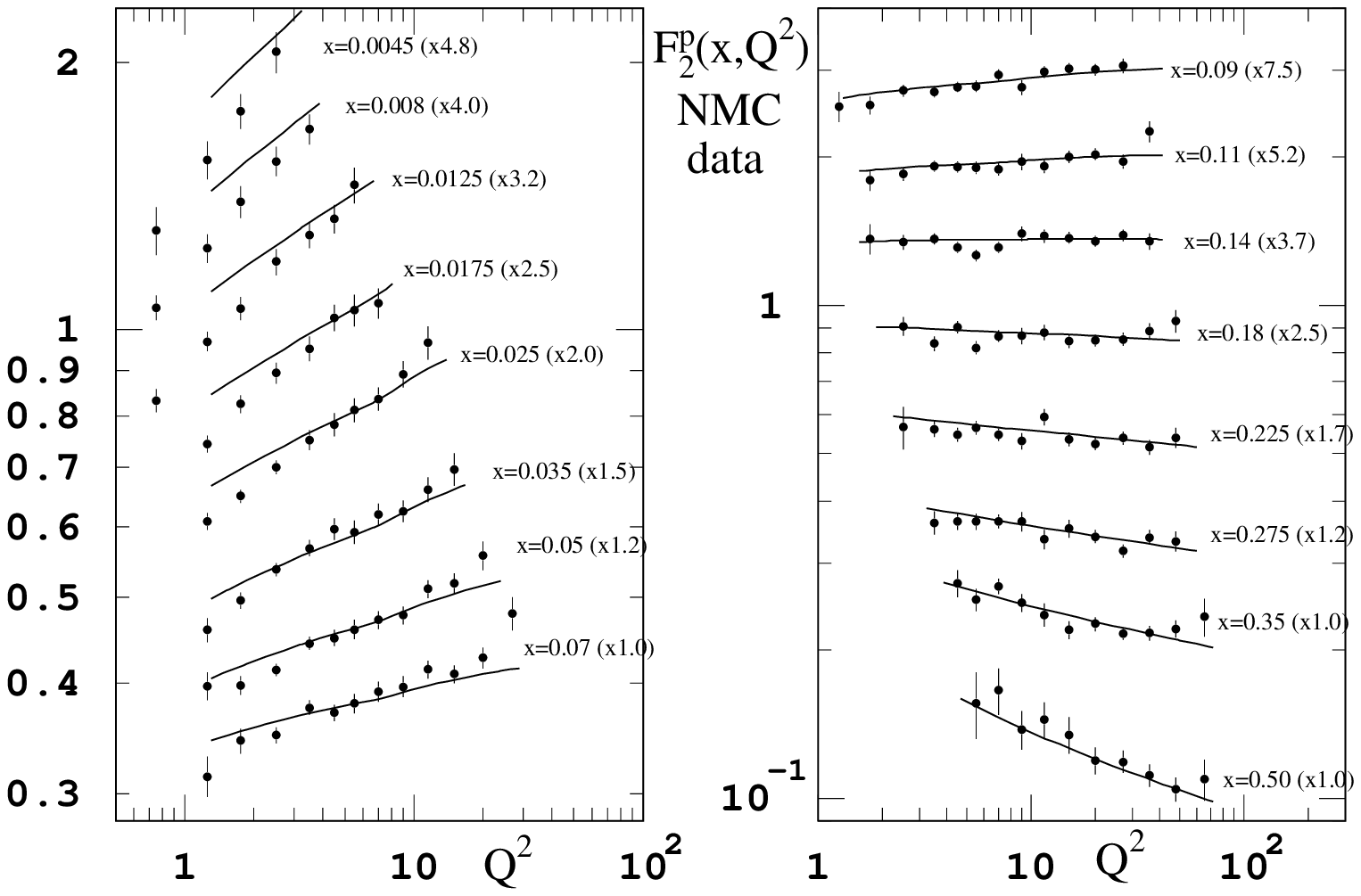,height=10.2in}}
\vspace{-11.5cm}
\caption{Comparison of the new fit using $\almz = 0.118$ with the
NMC data\protect\cite{nmc}.}
\label{nmcfig}
\end{figure}

\begin{figure}[htb] 
\centerline{\epsfig{file=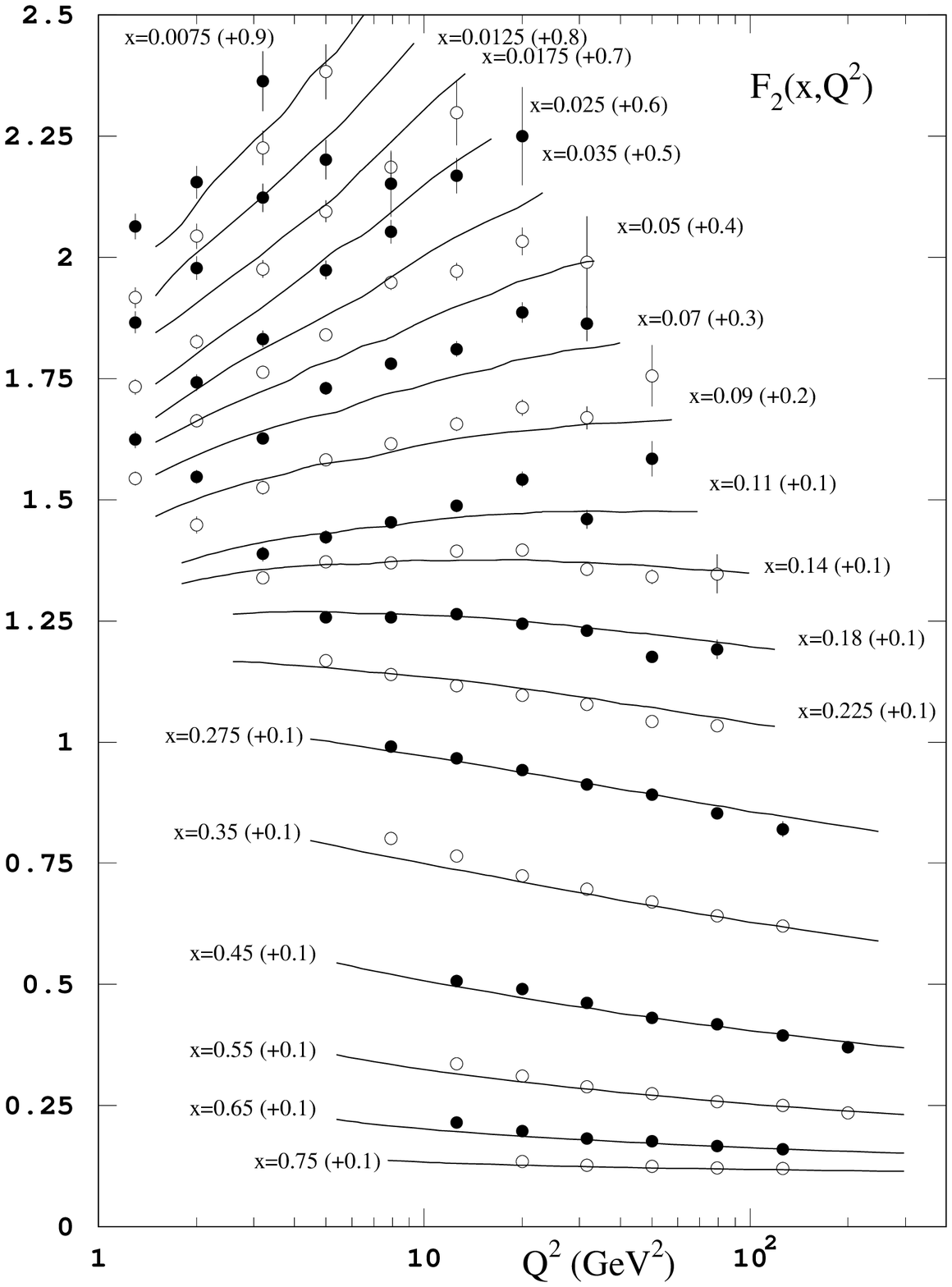,height=5.2in}}
\vspace{10pt}
\caption{Comparison of the new fit using $\almz = 0.118$ with the
new CCFR\protect\cite{ccfr} data on $F_2$. 
Here the theory curves have been multiplied
by a heavy target $x$ dependent correction factor and only the statistical
errors are shown on the data.}
\label{ccfr2fig}
\end{figure}

The resulting best value of $\almz$ which emerges from this overall
global analysis is 0.118 and we show comparisons of the fit 
using this value with a selection of data. The small $x$ data comparison
is shown in fig. \ref{smallx}.
The comparsion with the new NMC data is shown in fig. \ref{nmcfig}.
Notice that points below $Q^2= 2$ GeV$^2$, while shown in the figure, were
not included in the fit.
Finally we show the comparison with the new data from CCFR\cite{ccfr}
on $F_2$ extracted from $\nu$ and $\bar \nu$ interactions 
off an iron target in Fig. \ref{ccfr2fig}.
The failure to agree below $x$ of around 0.1 is well known. Basically these
low $x$ data prefer not to have the shadowing corrections, observed in muon
DIS, applied. Because of this conflict (with the NMC data for example) CCFR
$F_2$ below $x=0.1$ is dropped in the fit. Comparing the curves and large 
$x$ data in fig. \ref{ccfr2fig} one can see that the data prefer slightly
stronger $Q^2$ variation than that from using $\almz = 0.118$. Much
better consistency is found for $xF_3$ across the entire $Q^2$, $x$ range.

\begin{figure}[htb] 
\centerline{\epsfig{file=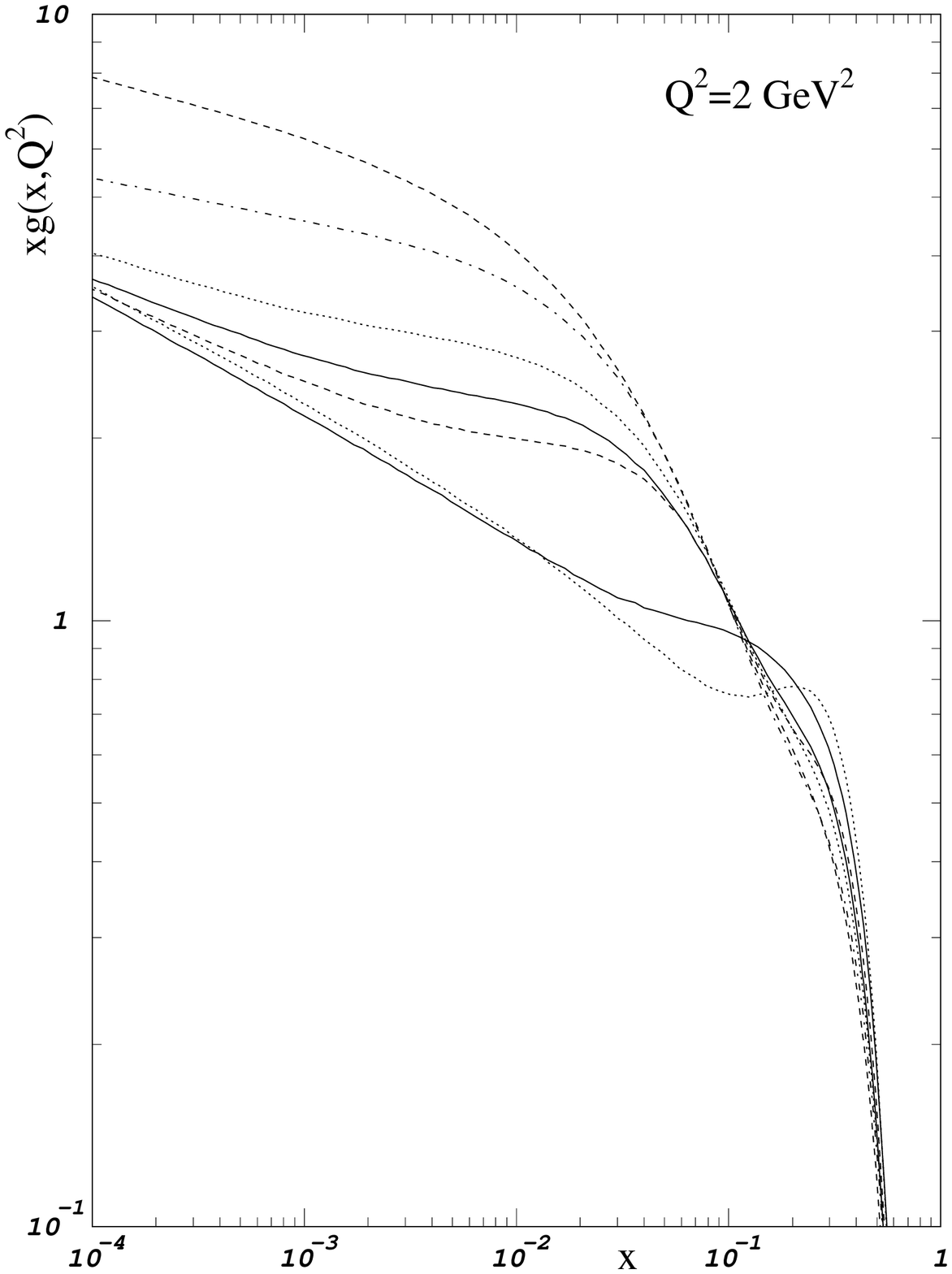,height=4.0in}}
\vspace{10pt}
\caption{Gluon distribution functions at $Q^2 = 2$ GeV$^2$ as $\almz$ is
varied from 0.105 (upper dashed) through 0.110 (dot-dashed), 0.115 (upper
dotted), 0.1175 (upper solid), 0.120 (lower dashed), 0.125 (lower solid)
and 0.1275 (lower dotted).}
\label{gluplot}
\end{figure}

In this analysis, the pdf's at $Q_0^2$ are parametrised in the usual 
MRS\cite{mrsr} way with $xg(x,Q_0^2) \sim x^{-\lambda_g}$ and
$xS(x,Q_0^2) \sim x^{-\lambda_s}$ for the glue and sea as $x \rightarrow 0$.
The exponents $\lambda_g$ and $\lambda_s$ are not constrained and as in
the previous analysis one finds that at such a low choice of $Q_0^2 =1$
GeV$^2$ $\lambda_s$ is positive (singular) while $\lambda_g$ comes out
negative $-$ i.e `valence-like' gluon. The actual value of the exponent
varies for different values of $\almz$ ($-0.15$ to $-0.21$) but when
$Q^2$ reaches just 2 GeV$^2$ all the gluon pdf's are approximately `flat',
see fig. \ref{gluplot}.
Taking the fit with $\alpha_s = 0.118$ and evolving to $Q^2 = 25$ GeV$^2$
yields pdf's which are extremely close to those shown in fig. \ref{figpdf}
so the different treatments of charm partons in the latest analyses of CTEQ 
and MRS actually lead to little numerical difference in the resulting pdf's.

\section*{Theoretical discussion of heavy quark production in DIS}
Let us consider possible approaches to treating the charm production
contribution to the structure function $F_2$.
\subsection*{Massless parton evolution}
The most simplistic
approach  is to assume that a probe of virtuality $Q^2$ can
resolve a charm quark pair in the proton sea when $Q^2 \gapproxeq
m_c^2$. Since such pairs originate from
fluctuations of the gluon field, $g \to c\overline{c}$, a
perturbative treatment should be valid as long as $m_c^2 \gg
\Lambda_{QCD}^2$.
As $Q^2$ increases, ${\cal O}(m_c^2/Q^2)$ corrections to the standard 
GLAP evolution become less important, and the charm quark  can 
be treated as a (fourth) massless quark. These ideas are embodied
in the `massless parton evolution' (MPE) approach
\bea
\label{eq:a1}
c(x,Q^2) &=& 0 \quad \mbox{for}\quad  Q^2 \leq \mu_c^2 \; ,
\nonumber \\
n_f &=& 3 + \theta(Q^2 - \mu_c^2) \quad \mbox{in}\quad P_{qg}, \;
P_{gg}, \; \beta_0, \ldots \; ,
\label{eq:nf}
\eea
where $\mu_c = {\cal O}(m_c)$. The charm contribution to the structure
function is then
\be
F_2^c(x,Q^2) = \textstyle{\frac{8}{9}} x c(x,Q^2)
\label{eq:a2}
\ee
in lowest order.  This is the approach adopted at NLO in the MRS
and CTEQ global parton analyses, with $\mu_c$ chosen in the MRS
analysis to achieve
a satisfactory description of the EMC $F_2^c$ data
\cite{emccharm}, while in the CTEQ analysis  $\mu_c$ is set equal
to $m_c$. 
For example, in the MRS(A) analysis \cite{mrsa}
it was found that $\mu_c^2 = 2.7$ GeV$^2$ and that this was to a
good approximation equivalent to taking
\be
2 c (x, Q_0^2) \; = \; \delta {\cal S} (x, Q_0^2)
\label{eq:a3}
\ee
with $\delta \approx 0.02$ at the input scale $Q_0^2 = 4$
GeV$^2$.  That is at the input scale, charm $(c + \overline{c})$
was found to have approximately the same shape as the total quark
sea distribution ${\cal S}$, and moreover to form about 2\% of
its magnitude.  

In NLO the charm structure function is given by
\be
F_2^c (x, Q^2) \; = \; \frac{8}{9} \: \int_x^1 \: dz \:
\frac{x}{z} \: \left [ C_{q = c} (z, Q^2, \mu^2) \: c \left (
\frac{x
}{z}, \mu^2 \right ) \: + \: C_g (z, Q^2, \mu^2) \: g
\left (\frac{x}{z}, \mu^2 \right ) \right ]
\label{eq:c1}
\ee
The charm quark coefficient function in (\ref{eq:c1}) has the form
\be
C_c \; = \; C_c^{(0)} \: + \: \frac{\alpha_S}{4 \pi} \: C_c^{(1)}
\: + \: \ldots,
\label{eq:c2}
\ee
while for the gluon we have
\be
C_g \; = \; \frac{\alpha_S}{4\pi} \: C_g^{(1)} \: + \: \ldots \:
\label{eq:c3}
\ee
and this expression with the standard massless $\msb$ coefficient functions
was used to calculate $F_2^c$ in the MPE approach.
\subsection*{Boson-gluon fusion}
A procedure which takes a very different approach is to generate $F_2^c$
from photon-gluon fusion (PGF) \cite{ghr,grs} alone. In this case, the
charm pdf appearing in eq.(\ref{eq:c1}) is identically zero everywhere
but the gluon coefficient function involves the non-zero charm mass $m_c$
and is the well-known PGF cross section, 
i.e. $C_g^{1}(z) \rightarrow C_g^{\rm PGF} (z,Q^2)$ with  
\bea
C_g^{\rm PGF} (z, Q^2) = \left \{ \left [ z^2 \: + \: (1 -
z)^2 \: + \: \frac{4 m_c^2}{Q^2} \: z (1 - 3z) \: - \:
\frac{8m_c^4}{Q^4} \: z^2 \right ] \right . \: \ln \: \frac{1 +
\beta}{1 - \beta} \nonumber \\
 \nonumber \\
 + \; \left . \left [ 8z (1 - z) \: - \: 1 \: - \:
\frac{4m_c^2}{Q^2} \: z (1 - z) \right ] \: \beta \right \} \;
\Theta \left ( Q^2 \: \left ( \frac{1}{z} - 1 \right ) \: - \:
4m_c^2 \right ).
\label{eq:pgf}
\eea
where $\beta$ is the velocity of one of the charm quarks in the
photon-gluon centre-of-mass frame
\be
\beta^2 \; = \; 1 \: - \: \frac{4m_c^2 \: z}{Q^2 (1 - z)}.
\label{eq:beta}
\ee
The $\Theta$ function in eq.(\ref{eq:pgf}), $\Theta (W^2 - 4m_c^2)$,
represents the $c\overline{c}$ production threshold, where $W$ is
the c.m.\ energy.  Its presence guarantees $\beta^2 \geq 0$.
The exact
next-to-leading order corrections to the PGF structure
function are known \cite{lvn} and  
recently \cite{bmsvn1} an analysis
has been used  to perform   $(\alpha_S \ln (Q^2/m_c^2))^n$
resummation up to ${\cal O}(\frac{m_c^2}{Q^2})$
for $Q^2 \gg m_c^2$. Such an approach
is of course not applicable in the threshold region
$Q^2 \gapproxeq   m_c^2$.
However for many purposes the PGF procedure is insufficient. If we consider
the high $Q^2$ limit of $C_g^{PGF}$ we get
\be
C_g^{\rm PGF} \; \rightarrow \; \left \{ z^2 \: + \: (1 - z)^2
\right \} \; \ln \left( \frac{1-z}{z}\; \frac{Q^2}{m_c^2}
\right) \: + \: 8z (1 - z) - 1
\label{eq:c16}
\ee
as $m_c \rightarrow 0$, which differs from the exact $m_c = 0$
coefficient $C_g^{(1)}$ by the presence of $Q^2/m_c^2$ in the
argument of the logarithm. This is a signal of potentially large
logs which should be resummed by GLAP evolution. Thus at large $Q^2$
we should include the charm quark as a parton in GLAP evolution.
In fact we should be able to provide a charm pdf all the way down
to the charm threshold in order to insert all relevant pdf's into
calculations of other perturbative QCD processes. The goal therefore
is to do this in a consistent way
\subsection*{Variable flavour number scheme procedure}
Here the procedure is to have no charm pdf below $Q^2=m_c^2$ 
but for $Q^2\geq m_c^2$ to have $c+\bar{c} \sim \alpha_s P_{qg} 
\ln Q^2/m_c^2$, 
where $P_{qg}$ is the splitting function $g\rightarrow c\bar{c}$ 
evaluated with $m=0$.  
Below $Q^2=m_c^2$ charm is produced solely by boson-gluon fusion as above and 
this is labelled the `fixed flavour number' scheme (FFNS).  Thus for $n_f =3$ we 
have an OPE for the structure functions
\be
F=C_{FFNS} \left(\frac{Q^2}{\mu^2} ,  \frac{m_c^2}{\mu^2} , 
\alpha_s (\mu^2) \right) \;
\hat{O}_{FFNS} (\alpha_s (\mu^2)) \; + \;{\cal O}(\frac{\Lambda^2}{Q^2})
\label{eq:opeffns}
\ee
Above $Q^2=m_c^2$ we work in the `variable flavour number' scheme (VFNS) 
with $n_f=4$ and write the OPE
\be
F=C_{VFNS} \left(\frac{Q^2}{\mu^2} ,   
\alpha_s (\mu^2) \right) \; 
\hat{O}_{VFNS} (\frac{Q^2}{\mu^2},\frac{m_c^2}{\mu^2} ,
\alpha_s(\mu^2) ) \; + \;
{\cal O}(\frac{m_c^2}{Q^2})
\label{eq:opevfns}
\ee
While the operators $\hat{O}_{VFNS}$ nominally depend on $m_c^2$ they can be 
chosen so that they {\it evolve} according to $m=0$ RGE's.  Thus in the VFNS 
the $m_c^2$ dependence appears only in the 
{\it corrections}. The coefficients $C_{VFNS}$ are simply the usual 
massless $\msb$ coefficients $C_{\msb}$.
The two schemes must be mutually 
consistent at all $Q^2 > m_c^2$ which implies 
(from eqns.(\ref{eq:opeffns},\ref{eq:opevfns})) that the 
${\cal O}(\frac{m_c^2}{Q^2})$ contribution in eq.(\ref{eq:opevfns}) can be 
written in the form 
\be
C'(\frac{m_c^2}{\mu^2})\; \hat {O}_{FFNS} (\alpha_s (\mu^2)) \;\; + \;\; 
{\cal O}(\frac{\Lambda^2}{\mu^2})
\label{eq:js1}
\ee
where $C'$ is a coefficient which $\sim$ powers of $\frac{m_c^2}{\mu^2}$. 
If we express the relation between the 
operators of the two schemes through a matrix $A$, 
\be
\hat{O}_{VFNS} = A\left (\frac{m_c^2}{\mu^2} , \alpha_s(\mu^2)\right ) \; 
\hat{O}_{FFNS}(\alpha_s)
\label{eq:js2}
\ee
then, in particular, the charm pdf is determined purely in terms of the 
{\it light} parton pdf's $-$ principally the gluon distribution of course.  
The structure functions 
above $Q^2=m_c^2$ can again be expressed in terms of the 
VFNS operators but now 
up to  ${\cal O} (\frac{\Lambda^2}{Q^2})$ corrections,
\bea
F &=& \left [C_{VFNS} \left(\frac{Q^2}{\mu^2}, \alpha_s \right ) 
+ C'(\frac{m_c^2}{\mu^2}) A^{-1}\right] \;
\hat{O}_{VFNS} \; + \; {\cal O}(\frac{\Lambda^2}{Q^2}) \nonumber \\
 &=& C_{ACOT} \left(\frac{Q^2}{\mu^2}, \frac{m_c^2}{\mu^2},
\alpha_s \right ) 
\hat{O}_{VFNS} \; + \; {\cal O}(\frac{\Lambda^2}{Q^2})
\label{eq:vfns1}
\eea
where $C_{ACOT}$ is the coefficient in the ACOT\cite{acot,wkt} 
formalism which then depends
on $m_c^2$. Thus in the ACOT formalism, while the partons obey a simple
evolution (independent of $m_c^2$), the coefficients would have a 
complicated dependence on $m_c^2$.
From eqns (\ref{eq:opeffns},\ref{eq:vfns1}) we obtain the relation between the 
coefficient functions \footnote{I am grateful to Robert 
Thorne for clarifying my understanding of the connection between the FFNS and 
VFNS.}
\be
C_{ACOT} \left(\frac{Q^2}{\mu^2}, \frac{m_c^2}{\mu^2},
\alpha_s \right ) = C_{VFNS} \left(\frac{Q^2}{\mu^2}, \alpha_s \right ) 
+ C'(\frac{m_c^2}{\mu^2}) A^{-1} \; =
C_{FFNS}\left(\frac{Q^2}{\mu^2} ,\frac{m_c^2}{\mu^2}, \alpha_s\right) A^{-1} 
\label{eq:coeffs}
\ee
where the term $C'A^{-1}$ on the rhs $\rightarrow 0$ 
as $Q^2 \rightarrow  \infty$.  

Buza {\it et al.}\cite{bmsvn1} have evaluated the coefficient functions 
for $Q^2 \gg m_c^2$ in the VFNS and the FFNS to NLO.  They find that for $Q^2
\gapproxeq 20 {\rm GeV}^2$ the two schemes agree very closely. 
The same authors have worked out the
relation (\ref{eq:js2}) in the limit $Q^2 \gg m_c^2$ exactly through 
terms $\ln^2 (m_c^2/\mu^2)$, $\ln(m_c^2/\mu^2)$ and constants.  
This infers a matching of all the
four flavour pdf's above the scale $\mu^2 = m_c^2$, to
the three flavour pdf's below. Again this is only
for $Q^2 \gg m_c^2$, which does not solve the problem of what to do
at $Q^2 \sim m_c^2$.  So there has to be another matching between
the calculated $c \overline{c}$ production cross section in the
threshold region to the charm pdf prescription at $Q^2 \gg
m_c^2$.  In leading order this is exactly the ACOT procedure.  It
remains to be investigated how to do this retaining terms 
${\cal O}(m_c^2/Q^2)$ at NLO.

Thus although the ${\rm FFNS} \rightarrow {\rm VFNS}$ change of
factorization scheme is very attractive in principle, a major
problem arises when one tries to
go down in $Q^2$ (still at NLO) all the way to the charm threshold - 
retaining the $C'A^{-1}$  term $-$ which is the eventual aim of ACOT. 
The $A^{-1}$  term is the one which contains the subtraction in the 
gluon coefficient functions   
for $F^C_2$ in eq.(\ref{eq:a2}) to avoid double counting
which at LO is quite straightforward and takes the form
$$
C_g^{(1)}(z) = C^{PGF}_g (z) - \alpha_s C_c^{(0)} \times
P_{qg}^{(0)} \ln Q^2/m_c^2 
$$ 
The effect of this is that $F^c_2$ is 
described at low 
$Q^2$ by PGF but as $Q^2 \rightarrow \infty$ by the evolved charm pdf.

Since the way it has been formulated is 
only at {\it leading order} (the charm pdf evolves only 
via $P^{(0)}_{qg} (z)$) 
this allows the standard expression to be used 
for the gluon 
coefficient function.  We are still left with the problem of how to evolve
{\it all} the pdf's beyond L.O. - in the $\msb$ scheme say.  
While the splitting 
functions will continue to have the $m=0$ forms, the 
entire $m^2$ dependence 
is put into the coefficient functions which will make them (the light parton 
coefficient functions as well as the charm coefficient functions) extremely 
complicated when they are eventually computed.

\subsection*{MRRS procedure}

In view of the practical difficulties of implementing 
the ACOT procedure at NLO, 
Ryskin and MRS\cite{mrrs} introduce an alternative prescription 
which is consistent at all orders but which generates partons which 
may be used with the {\it conventional} $\msb$ coefficient functions.  
The charm quark evolution is 
modified however $-$ of course 
only for $Q^2 \gg m_c^2$ does it evolve according to the 
massless $\msb$.  The modified splitting function is derived by studying the 
leading log decomposition of the Feynman diagrams and takes the form 
\bea
P^{(0)}_{cg} (z,\frac{m_c^2}{Q^2}) &=& \left[P^{(0)}_{qg} (z) \mid_{m=0} + 
\frac{2m_c^2}{Q^2} z(1-z)\right] \;\Theta (Q^2-m_c^2)\nonumber \\
P^{(1)}_{cg} (z,\frac{m_c^2}{Q^2}) &=& P^{(1)}_{qg} (z) \mid_{m=0,\msb} 
\label{eq:mrrs1}
\eea

Thus as $Q^2 \gg m_c^2$, $P_{cg}(z,\frac{m_c^2}{Q^2}) \rightarrow P_{qg} (z) 
\mid_{m=0}$ in the $\msb$ scheme.  The NLO splitting function 
obtained from the ladder graph actually has a natural scale given by 
$Q^2 = (m_c^2+k^2_T)/z(1-z)$. 
This choice was not the one chosen when the $\msb$ scheme was set up;
instead the scale $Q^2 = (m_c^2+k^2_T)$ was chosen
and it is the latter scale which leads to the 
expression in eq.(\ref{eq:mrrs1}).   
The `natural' choice of scale would have the advantage that it
can be interpreted as 
corresponding to a `resolution threshold' of $Q^2=4m_c^2$ below which 
$Q^2$ is too small to allow sufficient time to observe the 
$g\rightarrow c\bar{c}$ fluctuations.  
To avoid double counting the gluon coefficient function for 
$F^c_2$ must have a subtraction term, i.e.
\be
C^{(1)}_g  (z,\frac{m_c^2}{Q^2})  = C^{PGF}_g  (z,\frac{m_c^2}{Q^2})  
- \Delta C_g  
(z,\frac{m_c^2}{Q^2}) 
\label{eq:mrrs2}
\ee
where
\be
\Delta C_g  (z,\frac{m_c^2}{Q^2})  \sim \int^{Q^2}_{m_c^2} P^{(0)}_{cg} 
(z,\frac{m_c^2}{Q^2}) \; {\rm d} (\ln Q^2)
\label{eq:mrrs3}
\ee

\begin{figure}[htb] 
\centerline{\epsfig{file=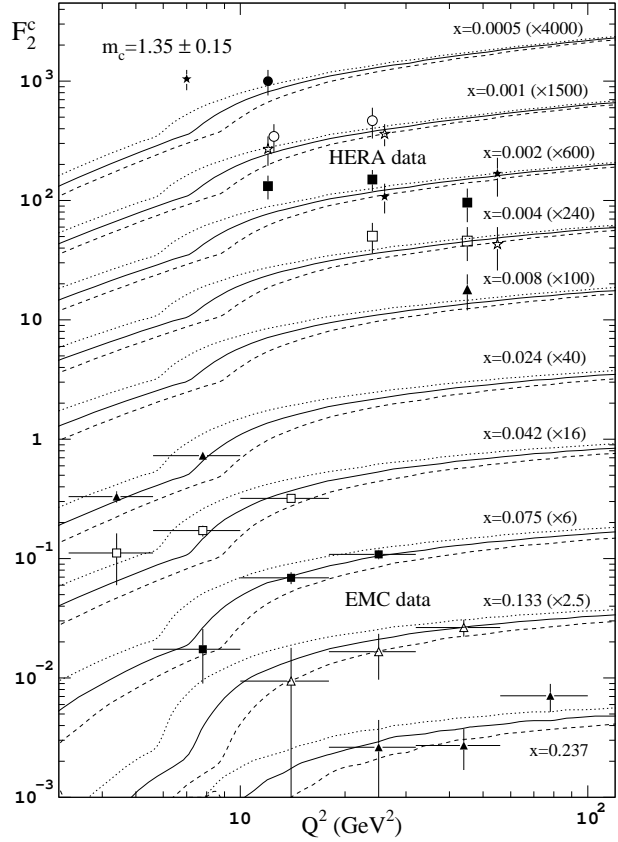,height=5.2in}}
\vspace{10pt}
\caption{Comparison of the data on the charm structure function with the
analysis of ref.\protect\cite{mrrs}. The HERA data are from 
refs\protect\cite{charmh1,charmzeus}, the Zeus points denoted by stars.
The large $x$ data are from EMC \protect\cite{emccharm}.} 
\label{figmrrs}
\end{figure}

This subtraction ensures the required cancellation of the leading 
contribution to 
the charm quark evolution around $Q^2 \sim m_c^2$, but the NLO 
contribution from the $P^{(1)}_{cg}$ term is still significant.  
However inserting the condition that the natural scale should really 
be $Q^2\sim 4m_c^2$ into the charm quark coefficient function forces 
the low $Q^2$ description of $F^c_2$ to 
be given entirely by the PGF term.  The resulting description of the measured 
$F^c_2$ cross sections is remarkably successful - see fig. \ref{figmrrs}.
\section*{Large x}
\begin{figure}[htb] 
\centerline{\epsfig{file=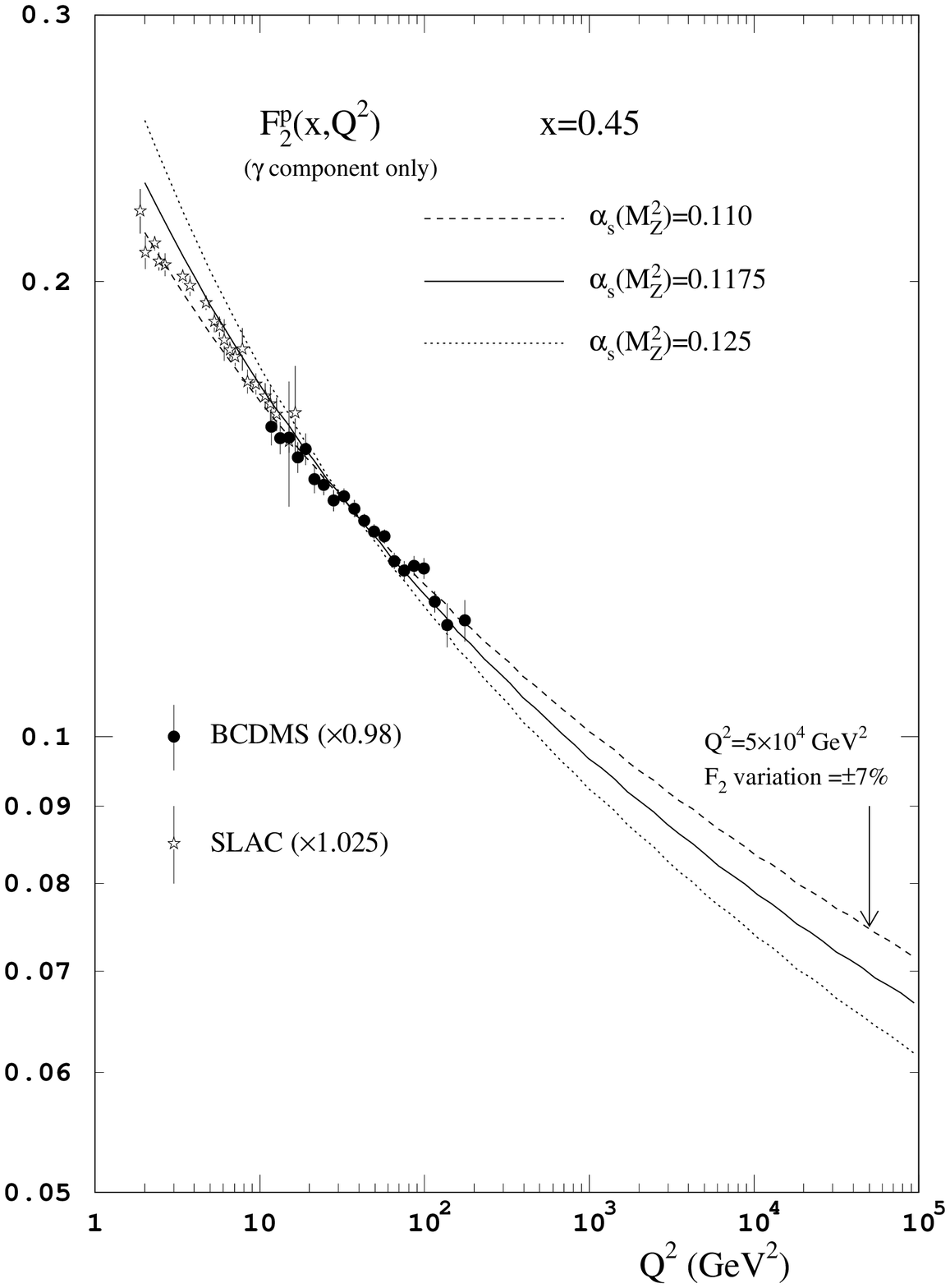,height=5.2in}}
\vspace{10pt}
\vspace{-1cm}
\caption{$F_2(e.m.)$ at $x=0.45$. The data shown are from BCDMS
\protect\cite{bcdms} and from SLAC\protect\cite{slac}.}
\label{figx45}
\end{figure}
Because of the topical interest in the magnitude of the cross section
at large $x$ and large $Q^2$
it is worthwhile to study the precision by which we believe
we understand the conventional DIS cross section. This translates into
estimating our confidence in the large $x$ pdf's. In fig. \ref{figx45}
we show the electromagnetic contribution to $F_2$ at $x=0.45$. For 
$Q^2 < 200$ GeV$^2$ the data have a normalisation uncertainty of
about 3$\%$. Taking the pdf's from the latest global analyses 
together with a generous uncertainty in the value of
$\almz$ $-$ from 0.110 to 0.125 we see that this gives an additional 
7$\%$ to the extrapolated value at $Q^2 \sim 10^5$ GeV$^2$. 
Including the standard interference term with the weak neutral
current the resulting cross section for $e^+ p \rightarrow e^+ X$
in the region of the observed excess at HERA
is then determined to within about 10$\%$.

\section*{Re-summed \protect{$\ln (1/x)$} terms}
So far I have discussed fits using LO and NLO resummation of large $\log Q^2$ 
terms.  Now we should ask is this really good enough when, at HERA, we 
explore $x$ below $10^{-5}$, i.e. $\ln 1/x > 9$, and the BFKL resummation of  
$\ln 1/x$ terms i.e. $\Sigma \alpha_s^n 
( \ln 1/x)^m$ should somehow be included.  
Catani and Hautmann\cite{ch1}, by incorporating $k_T-$factorisation into 
the collinear factorisation framework, calculated the relevant anomalous 
dimensions to lowest order in $\alpha_s$ for each power of $\ln (1/x)$ and 
similarly for the coefficient functions.  This prompted attempts to calculate 
structure functions within the conventional R.G. framework but including the 
leading $\ln (1/x)$ terms.  Comparisons with data suggested no improvement 
over fits with soft inputs and without $\ln (1/x)$ resummation \cite{bf2,frt} 
and indeed could worsen the agreement with data.  
Furthermore there was a strong 
dependence on the choice of factorisation scheme in which 
the leading $\ln 1/x$ terms were included\cite{bf2,frt,c2,c3}.

\begin{figure}[htb] 
\centerline{\epsfig{file=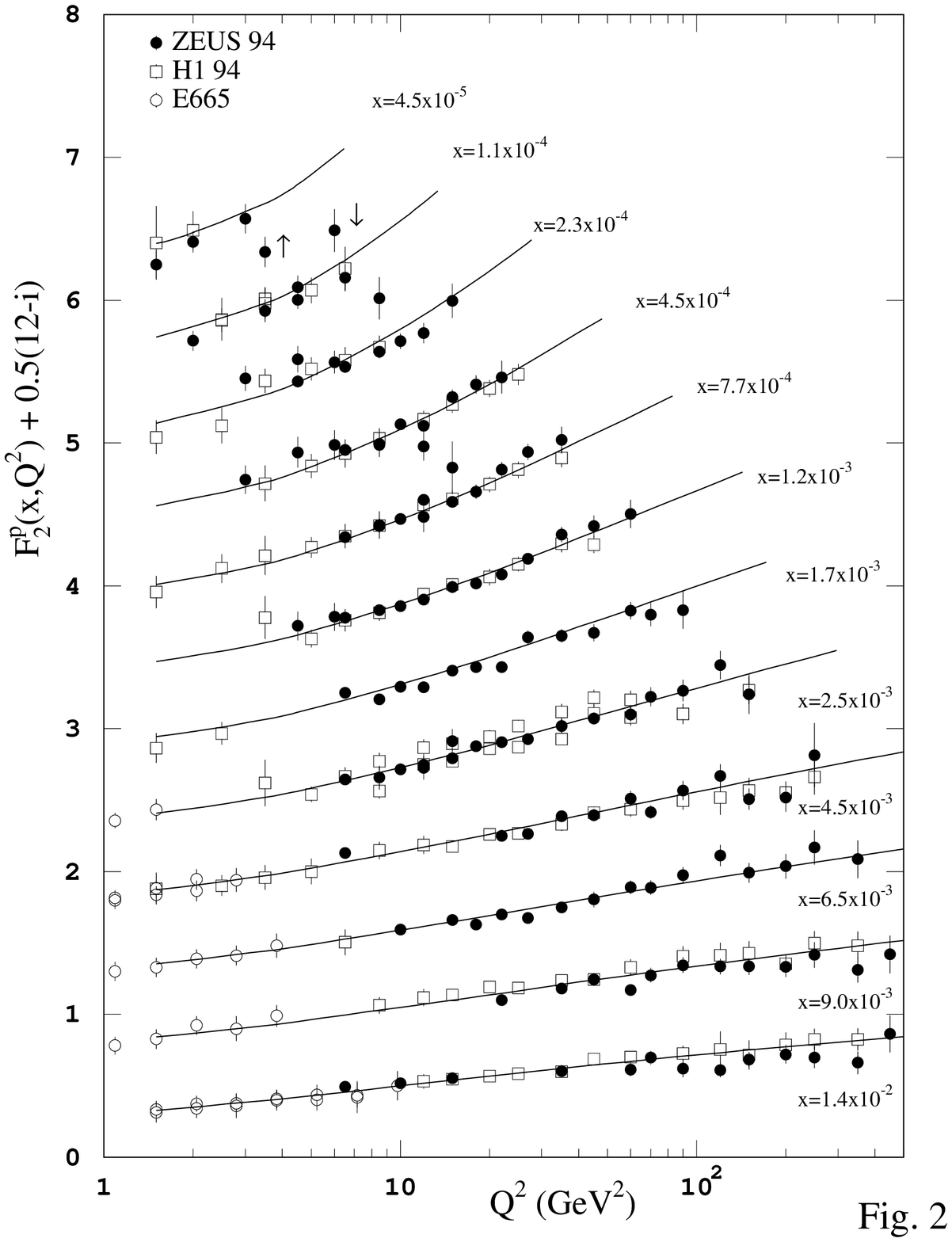,height=5.2in}}
\vspace{10pt}
\vspace{-1cm}
\caption{Comparison of the low $x$ data for $F_2(x,Q^2)$ with the leading
order renormalisation scheme consistent fit from ref\protect\cite{thorne}.}
\label{thornefig}
\end{figure}

The key to incorporating the resummation contributions is to derive an 
expansion for the structure functions which is leading order in 
both $\ln (1/x)$ {\it and } $\alpha_s(Q^2)$ and which is 
renormalisation scheme consistent.  
Thorne\cite{thorne} has shown that this automatically leads to 
results which are factorisation scheme invariant and provides a 
framework in which the `physical' 
anomalous dimensions of Catani\cite{c4} emerge naturally.  This framework is 
discussed more fully in Thorne's talk\cite{thorne2}.  There is a decided 
improvement in the description of the $F_2$ data as a result of including the 
$\ln (1/x)$ contributions in this theoretically consistent manner, 
especially at small $x$.  Fig.\ref{thornefig} shows a comparison 
with the low $x$ data and such a fit 
demonstrating the better agreement at very low $x$.

Perhaps the sharpest prediction of this approach is the behaviour of 
$F_L(x,Q^2)$ at small $x$ where the magnitude of the longitudinal 
structure function is much smaller than the `conventional' two loop 
prediction.  Indeed as a final note for 
this talk I would stress the importance of measuring directly $F_L$ at 
HERA as a probe of dynamics of small $x$ physics.

\section*{Acknowledgments}
I am grateful for many discussions with Alan Martin, Misha Ryskin,
James Stirling and Robert Thorne and for comments from Jack Smith.


\begin{references}
\bibitem{nmc} NMC collab.: M. Arneodo {\it et al.} hep-ph/9610231, to
be published in Nucl. Phys.
\bibitem{ccfr} CCFR collab.: W.G. Seligman {\it et al.} hep-ex/9701017,
to be published in Phys. Rev. Lett. and W.G. Seligman, Ph.D. thesis
(Columbia University), Nevis report 292.
\bibitem{charmh1} H1 collab.: C. Adloff {\it et al.}, DESY-96-138 (1996)
\bibitem{charmzeus} ZEUS collab.: J. Roldan {\it et al.}, talk in the
WG1 session at this workshop.
\bibitem{thorne} R.S. Thorne, Phys. Lett. {\bf B392} 463 (1997);
Rutherford Lab. report RAL-TR-96-065, hep-ph/9701241.
\bibitem{lai} H.L. Lai and W. K. Tung, CTEQ preprint MSU-HEP-61222,
CTEQ-622 (1997).
\bibitem{acot} M.A.G.\ Aivazis, J.C.\ Collins, F.I.\ Olness and
W.-K.\ Tung, Phys.\ Rev.\ {\bf D50} 3102 (1994).
\bibitem{h1f2}
H1 collaboration: S.~Aid {\it et al.}, Nucl. Phys.
{\bf B470} 3 (1996).
\bibitem{wkt} J.C.\ Collins and W.-K.\ Tung, {\it Nucl.\ Phys.}
{\bf B278} 934 (1986); F.I.\ Olness and W.-K.\ Tung, {\it Nucl.\
Phys.} {\bf B308} 813 (1988); M.A.G.\ Aivazis, F.I.\ Olness and
W.-K.\ Tung, {\it Phys.\ Rev.\ Lett.} {\bf 65} 2339 (1990).
\bibitem{mrrs} A. D. Martin, R. G. Roberts, M. G. Ryskin and W. J. 
Stirling, preprint DTP/96/102, RAL-TR-96-103.
\bibitem{mrsr} A.D.~Martin, R.G.~Roberts and
W.J.~Stirling, Phys. Lett. {\bf B387} 419 (1996).
\bibitem{zeusf2}
ZEUS collaboration: M.~Derrick {\it et al.}, Zeit. Phys. {\bf
C69} 607 (1996);  Zeit. Phys. {\bf C72} 399 (1996).
\bibitem{bf1} R. D. Ball and S. Forte, Phys. Lett. {\bf B358} 365 (1995).
\bibitem{bcdms} BCDMS collaboration:  A.C.~Benvenuti {\it et
al.}, Phys. Lett. {\bf B223} 485 (1989).
\bibitem{slac}
L.W.~Whitlow {\it et al.}, Phys. Lett. {\bf B282} 475 (1992);
L.W.~Whitlow, preprint SLAC-357 (1990). 
\bibitem{e665}
E665 collaboration: M.R.~Adams {\it et al.}, Phys. Rev. {\bf D54}
3006 (1996).
\bibitem{emccharm} EMC collab.: J.J.~Aubert {\it et al.}, 
Nucl. Phys. {\bf B213} 31 (1983).
\bibitem{mrsa} A.D.~Martin, R.G.~Roberts and 
W.J.~Stirling, Phys. Rev. {\bf D50} 6734 (1994).
\bibitem{ghr} M.\ Gl\"{u}ck, E.\ Hoffmann and E.\ Reya, Z.\
Phys.\ {\bf C13} 119 (1982).
\bibitem{grs} M.\ Gl\"{u}ck, E.\ Reya and M.\ Stratmann, Nucl.\
Phys.\ {\bf B422} 37 (1994).
\bibitem{lvn} E.\ Laernen, S.\ Riemersma, J.\ Smith and W.L.\ van
Neerven, Nucl.\ Phys.\ {\bf B392}  162 (1993).
\bibitem{bmsvn1} M.\ Buza, Y.\ Matiounine, J.\ Smith  and W.L.\
van Neerven, preprint NIKHEF/96-027 (1996).
\bibitem{bmsvn2} M.\ Buza, Y.\ Matiounine, J.\ Smith  and W.L.\
van Neerven, preprint DESY 96-258 (1996).
\bibitem{ch1} S. Catani and F. Hautmann, Phys. Lett. {\bf B315} 157
(1993); Nucl. Phys {\bf B427} 475 (1994).
\bibitem{bf2} R.D. Ball and S. Forte, Phys. Lett. {\bf B351} 313
(1995); Phys. Lett. {\bf B358} 365 (1995).
\bibitem{frt} J.R. Forshaw, R.G. Roberts and R.S. Thorne, Phys. Lett.
{\bf B356} 79 (1995).
\bibitem{c2} S. Catani, Zeit. Phys. {\bf C70} 263 (1996).
\bibitem{c3} M. Ciafaloni, Phys. Lett. {\bf B356} 74 (1995).
\bibitem{c4} S. Catani, talk at UK workshop on HERA physics, 1995,
unpublished; preprint hep-ph/9609263, DFF 248/4/96; preprint 
hep-ph/9608310, to appear in  Proc. DIS96. 
\bibitem{thorne2} R.S. Thorne, talk in the WG1 session at this workshop.

\end{references}
\end{document}